\documentclass[11pt,twoside]{article}
\usepackage{asp2010}

\aspvoltitle{Solar Polarization 7}
\aspvolauthor{K.~N. Nagendra, J.~O. Stenflo, Z. Qu and M. Sampoorna  eds.}
\aspcpryear{2013}

\resetcounters

\begin{document}
\label{deltaspot}

\allowtitlefootnote

\title{The magnetic configuration of a $\delta$-spot}
\author{H. Balthasar$^1$, C. Beck$^2$, R.E. Louis$^1$, M. Verma$^{1,3}$ and C. Denker$^1$}
\affil{$^1$Leibniz-Institut f\"ur Astrophysik Potsdam, An der Sternwarte 16, 14482 Potsdam,
      Germany}
\affil{$^2$National Solar Observatory, Sacramento Peak, 3010 Coronal Loop, 
         Sunspot New Mexico 88349, U.S.A.}
\affil{$^3$Max-Planck-Institut f\"ur Sonnensystemforschung, Max-Planck-Stra\ss e 2, 
       37191 Katlenburg-Lindau, Germany}

\begin{abstract}
Sunspots, which harbor both magnetic polarities within one
penumbra, are called $\delta$-spots. They are often associated
with flares. Nevertheless, there are only very few
detailed observations of the spatially resolved magnetic field
configuration. We present an investigation performed
with the Tenerife Infrared Polarimeter at the Vacuum Tower
Telescope in Tenerife. We observed a sunspot with a main umbra
and several additional umbral cores, one of them with opposite magnetic 
polarity (the $\delta$-umbra). 
The $\delta$-spot is divided into two parts by a line along which 
central emissions of the spectral line Ca\,{\sc ii}\,854.2\,nm appear.
The Evershed flow comming from the main umbra ends at this line.
In deep photospheric layers,
we find an almost vertical magnetic field for the $\delta$-umbra,
and the magnetic field decreases rapidly with height, faster than 
in the main umbra.
The horizontal magnetic field in the direction connecting main
and $\delta$-umbra is rather smooth, but in one location next to a 
bright penumbral feature at some distance to the $\delta$-umbra, we
encounter a change of the magnetic azimuth by 90$^\circ$ from one 
pixel to the next. Near the $\delta$-umbra, but just outside,
we encounter a blue-shift of the spectral line profiles which we 
interpret as Evershed flow away from the $\delta$-umbra.
Significant electric current densities are observed at the dividing 
line of the spot and inside the $\delta$-umbra.
\end{abstract}

\section{Introduction}
\label{balthasar-sec-intro}
Normally, the two magnetic polarities appear in two or more separated sunspots.
\citet{balthasar-kuenzel1} described cases where both polarities appear within a 
single penumbra,
and he named them $\delta$-sunspot groups. They are frequently associated 
with flares \citep[see][]{balthasar-sammis}. \citet{balthasar-kuenzel2} later 
suggested to extend the Hale classification of sunspots to include $\delta$-spots.

Flares can be ignited when shear flows along the Polarity Inversion 
Line (PIL) build up magnetic shear or twist. Therefore, velocities in and around 
$\delta$-spots have been investigated frequently. \citet{balthasar-tan} observed a 
shear flow of 0.6\,km\,s$^{-1}$ between two umbrae of opposite polarity before
an X3.4 flare. After the flare, the flow was reduced to 0.3\,km\,s$^{-1}$.
\citet{balthasar-denker} investigated a case where a shear flow did not change 
the magnetic shear sufficiently and concluded that a shear flow might even reduce 
magnetic shear. \citet{balthasar-lites} found flows converging at the PIL.
Remarkable downflows near the PIL have been reported by \citet{balthasar-valentin},
who found 14\,km\,s$^{-1}$, while \citet{balthasar-takizawa} detected values between 1.5 
and 1.7\,km\,s$^{-1}$. Doppler velocities of $\pm$10\,km\,s$^{-1}$ were detected by 
\citet{balthasar-fischer} during flaring activity.
\citet{balthasar-minchae} observed that the opposite polarity part of a sunspot rotated  
around its center by 540$^\circ$ within five days. Such rotation can cause coronal
mass ejections as demonstrated by \citet{balthasar-tibor}.
\citet{balthasar-wang} reported a formation of a penumbra between dark features of 
opposite polarity associated with a C7.4 flare. After the flare, a new $\delta$-spot 
was created.
\citet{balthasar-jennings} used the Mg\,{\sc i} line at 12.32\,$\mu$m to observe a large sunspot group just prior to the occurrance of an M2-flare. The flare was initialized by
flux cancellation at a location where opposite polarities were close together.

In this work, we observed a $\delta$-sunspot to investigate its detailed magnetic 
structure and to search for special magnetic 
configurations that might lead to flares. Parts of the results are published in 
\citet{balthasar-aaletter}. 

\section{Observations and data reduction}
\label{balthasar-sec-observation}

\begin{figure}[t]
\begin{center}
\includegraphics[width=132mm]{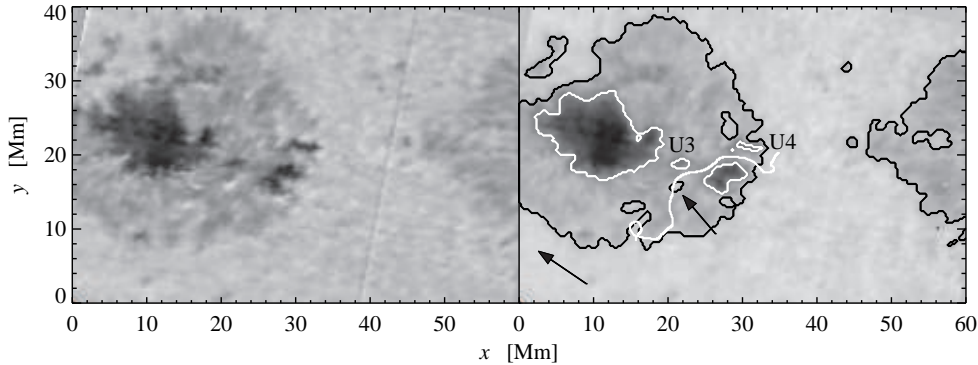}
\caption{Slit reconstructed continuum images of the spot near 
Fe\,{\sc i}\,1078.3\,nm (left), and near Ca\,{\sc ii}\,854.2\,nm (right). 
The arrow in the lower left corner points to disk center. 
Contours mark the boundaries of umbra and penumbra. 
Small umbrae are marked as U3 and U4, and another arrow points to a bright
feature close to the PIL (thick white line).
The axis orientation corresponds to the solar North and West directions.
}
\label{balthasar_kont}
\end{center}
\end{figure}

The sunspot group NOAA~11504 consisted on 2012, June 17 of three major spots, 
and one of them harbored smaller umbrae other than the main umbra. 
Close to the outer boundary but still inside
the complex penumbra was an extended dark feature with opposite magnetic polarity
with respect to the main umbra. In the  following, we call this feature the 
$\delta$-umbra. The group was located 35$^\circ$ from disk center at 
18$^\circ$S/29$^\circ$W (cos $\vartheta$ = 0.82).
We observed this sunspot with the Vacuum Tower Telescope (VTT) in Tenerife.
The full Stokes vector in two different spectral lines, Fe\,{\sc i} 1078.3\,nm
and Si\,{\sc i}\,1078.6\,nm, was recorded with the Tenerife Infrared Polarimeter 
\citep[TIP;][]{balthasar-tip}. 
As \citet{balthasar-hobagoem} pointed out, these two lines originate 
from different atmospheric layers except for the cool cores of umbrae.
Both lines form a normal Zeeman triplet with a splitting factor $g_\mathrm{eff} = 1.5$. 
The spectral dispersion was 2.19\,pm, and along the slit, two pixel were 
binned resulting in an image scale of 0\farcs35 per pixel.
During the time period 10:00 -- 10:38 UT we scanned the sunspot with 180 steps 
of 0\farcs 36 width \textbf{corresponding to the slit width}. Both spatial step widths correspond roughly to the theoretical 
resolution of the telescope of 0\farcs39.
For each scan position we accumulated ten exposures of 
250\,ms in each modulation state of TIP. Seeing influences were compensated by the 
Kiepenheuer Adaptive Optics System \citep[KAOS;][]{balthasar-kaos, balthasar-berkefeld}.
The same data set was also used by \citet{balthasar-aaletter}.

\begin{figure}[t]
\begin{center}
\includegraphics[width=132mm]{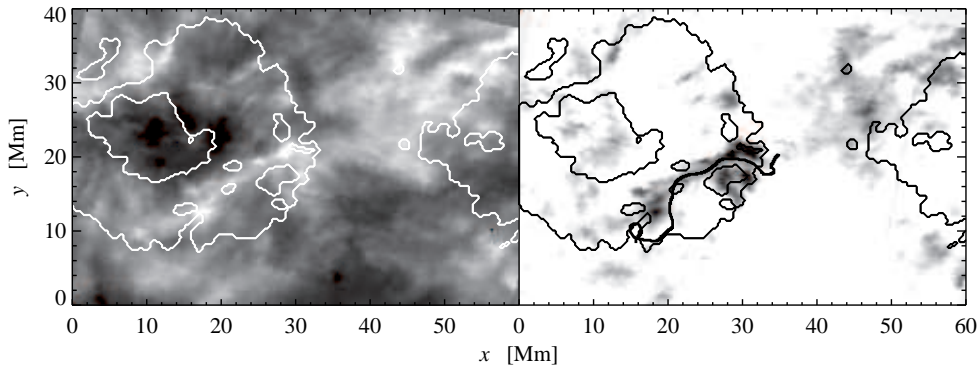}
\caption{Line core intensity of the line Ca\,{\sc ii}\,854.2\,nm (left)
and negative image of the central emission in this line (right).
Contours mark the boundaries of umbra and penumbra as well as the PIL. 
}
\label{balthasar_cair}
\end{center}
\end{figure}

The magnetic vector field and the Doppler velocities of these two lines were 
derived with the Stokes Inversion based on Response functions (SIR). This code
was developed by \citet{balthasar-sir}. We set three nodes for the temperature 
and kept magnetic field strength $B$, inclination $\gamma$, and azimuth $\psi$ 
constant with height, as well as the Doppler velocity $v_{\mathrm D}$. 
We obtained the height dependence of these parameters by
inverting the two lines separately. We used a single-component model atmosphere.
The SIR code provided also error estimates for the calculated quantities,
and we processed these errors by error propagation to get the errors of the final 
physical parameters.
The next issue was to solve the magnetic azimuth ambiguity. In the first step, we 
assumed a single azimuth center in the main umbra and chose that direction which
had the smaller difference to the radial orientation. It had to be considered 
that this orientation is inward because of the negative polarity on the main 
umbra. This way we obtained a roughly correct start azimuth (although the 
$\delta$-umbra had its own azimuth center) for the minimum energy method 
provided by \citet{balthasar-leka}. With this method, the term 
$|\nabla \cdot \vec B| + w|J_{\mathrm z}|$ is minimized. $J_{\mathrm z}$ is the 
vertical component of the electric current density and $w$ is a weighting factor.
The output delivers a reliable magnetic azimuth.
Finally we rotated all images to the local reference frame.

Next to TIP, we mounted a CCD-camera to record spectra of the line Ca\,{\sc ii}\,854.2\,nm
and another photospheric line Si\,{\sc i}\,853.6\,nm with a high excitation potential of 
6.15\,eV, which probes the deep layers of the solar atmosphere \citep[see][]
{balthasar-beckreza}. With this setup, we had a dispersion of 0.82\,pm per pixel.
This CCD-camera 
could not be synchronized with TIP. Therefore we integrated over 9\,s and used these data 
only as intensity spectra to determine Doppler velocities and line core intensities
of the Ca line. Doppler velocities of the Si line were determined from the minimum 
position of a parabola fit. Because this line becomes very weak in the cool parts of
the umbra, velocities are rather noisy there. For the Ca line, we distinguish between 
three different cases. If there was no central emission, as usual, we determined the 
minimum of a polynomial fit of fourth degree. If we detected central emission 
with a single peak,
we applied a parabola fit to this central emission and calculated its maximum position.
In some cases, we encountered a central reversal in the emission. In these cases, we 
fit a polynomial of fourth degree, and its central minimum was used to determine 
the Doppler velocity. Central emission occurs in the $\delta$-spot along a line 
dividing the lower right part of the penumbra in Fig.~\ref{balthasar_kont}. 
In the following, 
we call this line the `Central Emission Line' (CEL).  The CEL is partly co-spatial 
with the PIL, but not everywhere.
Fig.~\ref{balthasar_cair} shows the line core intensities of the infrared 
Ca line and the maximum intensity of the central emission after subtraction of a 
parabel fit through the wings of the line. 
%
\begin{figure}[t]
\begin{center}
\includegraphics[width=125mm]{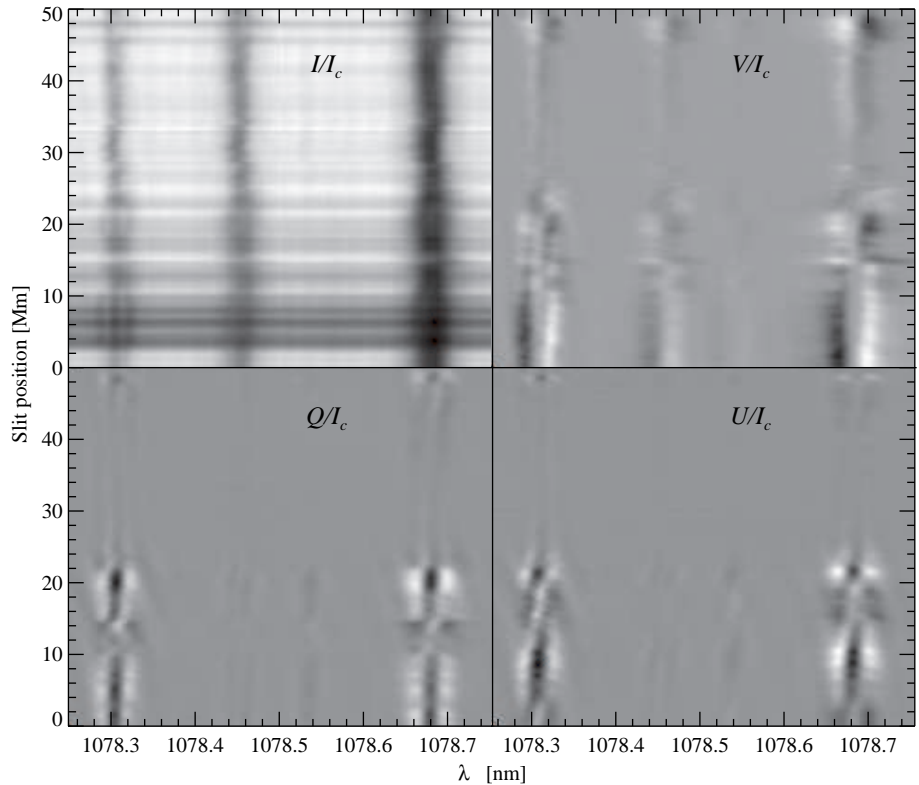}
\caption{Maps of the Stokes-profiles \textbf{$I/I_c$, $Q/I_c$, $U/I_c$, and $V/I_c$, 
normalized to the continuum intensity at disk center $I_c$} for a
selected slit position. 
\textbf{Scalings are: $I/I_c$ 0.35--1.0, $V/I_c \pm-0.1$, and $Q/I_c$ and 
$U/I_c \pm 0.08$.}
Note the weak multi-lobe $Q/I_c$-profiles at 12\,Mm along the slit.
}
\label{balthasar_stokes}
\end{center}
\end{figure}

All maps were destretched with the method described by \citet{balthasar-meetu} to get
quadratic pixels with a side length of 260\,km on the Sun.

\section{Results}
\label{balthasar-sec-results}

Slit reconstructed intensity maps  are shown in Fig.~\ref{balthasar_kont}.
The spot exhibited a complex structure of the penumbra, which harbored 
several dark features beside the main umbra. Most of them had the same 
polarity as the main umbra, but the bow-shaped feature in the South-West 
part of the penumbra was the $\delta$-umbra, i.e., it had the opposite 
polarity. The $\delta$-umbra had the same polarity as the leading spot of 
the group, which is partly seen on the right side of Fig.~\ref{balthasar_kont}.
We also see some bright inclusions in the penumbra, and one 
of them, marked by an arrow in Fig.~\ref{balthasar_kont}, seems 
to play a special role in the magnetic configuration. 
Between main and $\delta$-umbra there was a another small umbra U3, which 
had the same polarity as the main umbra.
Another small umbra U4 close to the $\delta$-umbra also had the polarity 
of the main umbra.

Maps of the Stokes-profiles at a selected slit position are
shown in Fig.~\ref{balthasar_stokes}. Weak multi-lobe $Q$-profiles appear at 12\,Mm
along the slit.

\begin{figure}
\includegraphics[height=132mm, angle=90]{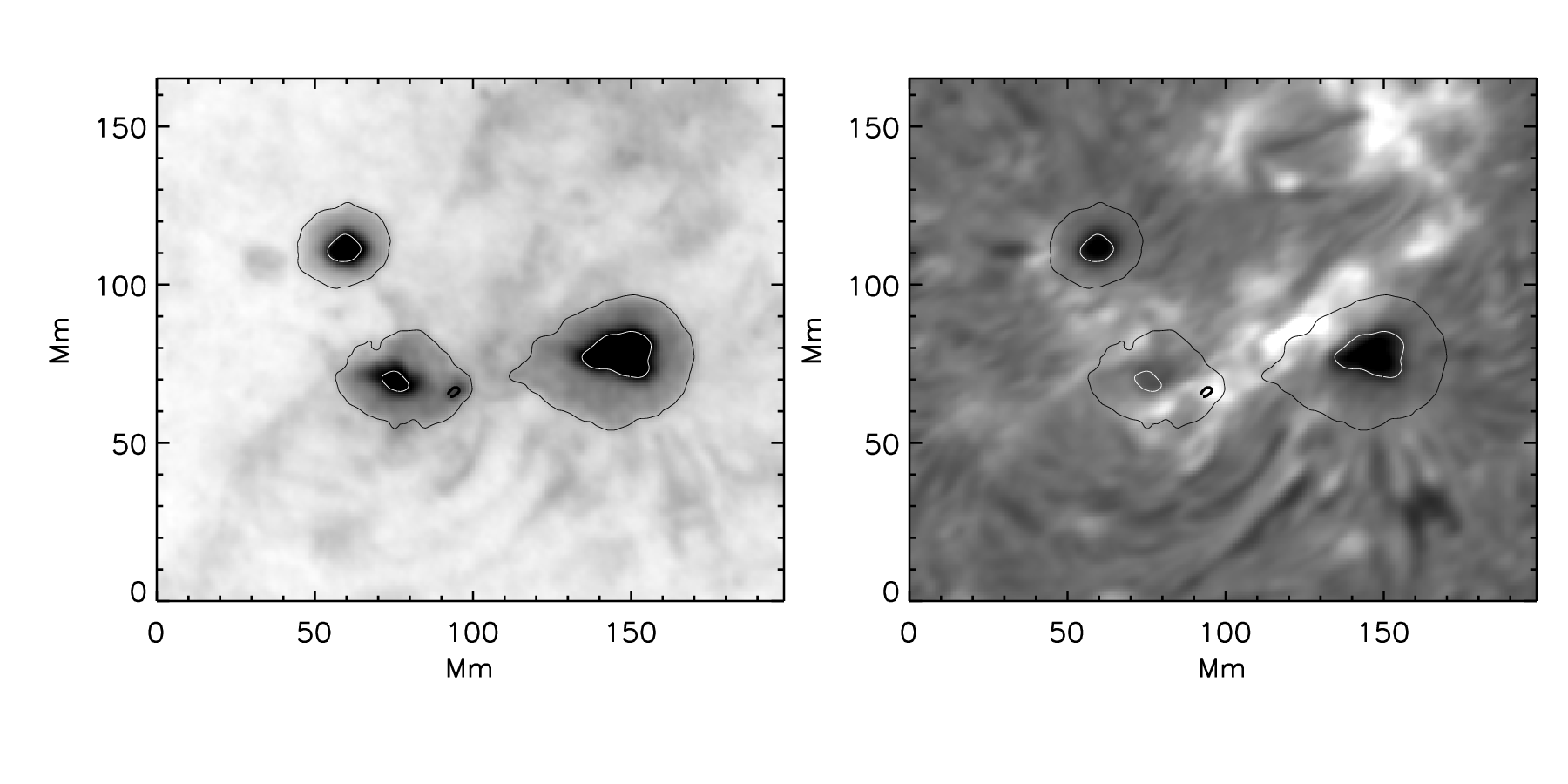}
\caption{Images obtained with ChroTel in the core of the
\ion{He}{i} line at 1083.03\,nm (left) and in H$\alpha$ (right). 
Black and white contours mark penumbra and umbrae, respectively.}
\label{balthasar_chrotel}
\end{figure}

Context images in the line cores of H$\alpha$ and the
\ion{He}{i} line at 1083.03\,nm were obtained with the 
Chromospheric Telescope \citep[ChroTel;][]{balthasar-bethge2011, balthasar-bethge2012}. 
On this day, the ChroTel observations started at 10:45\,UT, 
and we used the first images of the series
shown in Fig.~\ref{balthasar_chrotel}.
In the helium image, we see mainly
photospheric structures because this line is normally weak compared to H$\alpha$. 
In H$\alpha$, the umbrae
of the two neighboring spots ($x,y \sim$ 60\,Mm, 110\,Mm and
150\,Mm, 75\,Mm) are well visible,
while it is very hard to identify the umbrae of the $\delta$-spot ($x,y \sim$ 75\,Mm, 70\,Mm).
Here, the photospheric
structures were completely covered by chromospheric features indicating that
the magnetic field in the chromosphere
plays an important role in $\delta$-spots.
Especially, we detected a brightening in H$\alpha$ along the CEL. An even
stronger brightening occured in the prolongation of the CEL along the outer
penumbral boundary of the leading spot of this group.

\subsection{Magnetic field}
\label{balthasar-subsec-field}

\begin{figure}[t]
\begin{center}
\includegraphics[width=132mm]{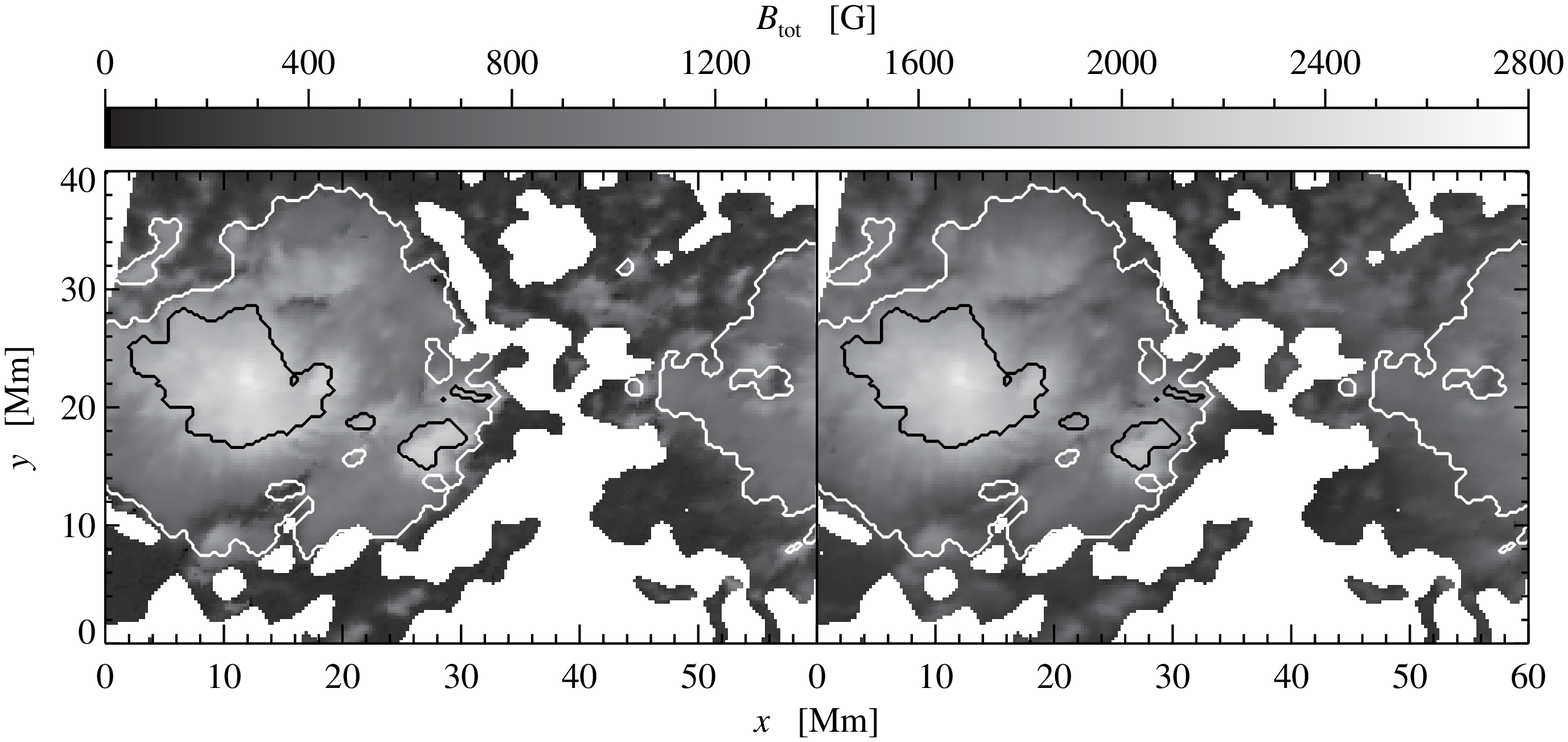}
\caption{Total magnetic field strength from Fe\,{\sc i}\,1078.3\,nm (left),
and from Si\,{\sc i}\,1078.6\,nm (right). 
Areas with insignificant polarization outside the spot are white.
}
\label{balthasar_btot}
\end{center}
\end{figure}

The total magnetic field strength is shown in Fig.~\ref{balthasar_btot}.
In the main umbra we found 2600\,G in the Fe\,{\sc i}\,1078.3\,nm line
and 2570\,G in the Si\,{\sc i}\,1078.6\,nm line, typical for medium sized sunspots.
The values in the $\delta$-umbra were 2250\,G and 2030\,G, respectively. 
The field strength dropped to 1500\,G from 
the iron line and and 1400\,G from the silicon line in the area between main 
and $\delta$-umbra. Umbral core U3 exhibited 2000\,G
in the iron line, but only 1500\,G in the silicon line. 

\begin{figure}[t]
\begin{center}
\includegraphics[width=132mm]{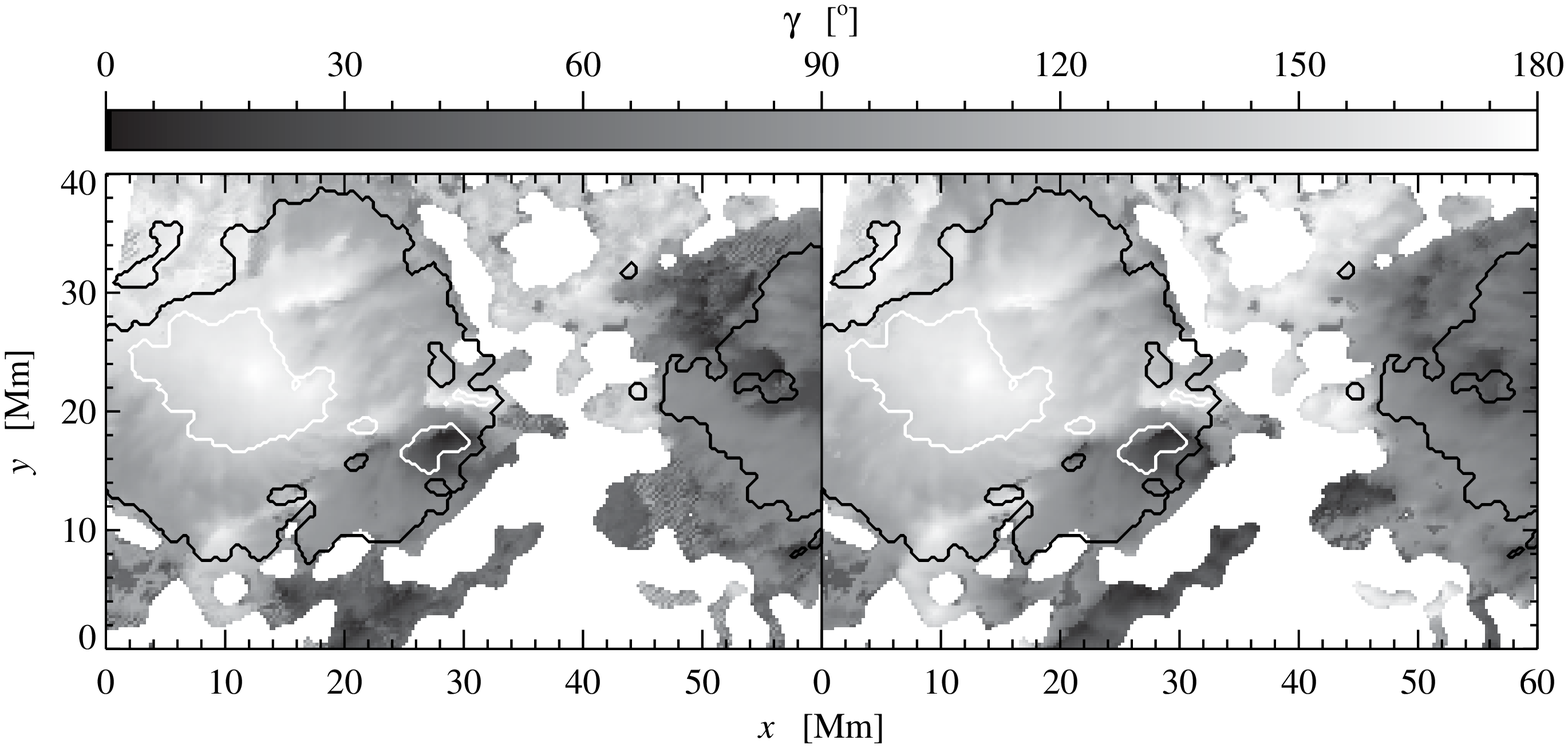}
\caption{Inclination of the magnetic field.
}
\label{balthasar_gamm}
\end{center}
\end{figure}

The magnetic inclination in Fig.~\ref{balthasar_gamm} was  close to 180$^\circ$
in the main umbra indicating the negative polarity. In and around the 
$\delta$-umbra, we encountered small inclinations because of the positive 
polarity here. Between the two umbrae the inclination changed rapidly from
about 150$^\circ$ to about 60$^\circ$, and only at the PIL it was horizontal
(90$^\circ$). Inside U3 and U4, the field was less inclined
than in its surroundings. The magnetic field was also more vertical 
at the CEL, where it was not cospatial with the PIL.

\begin{figure}[!ht]
\begin{center}
\includegraphics[width=132mm]{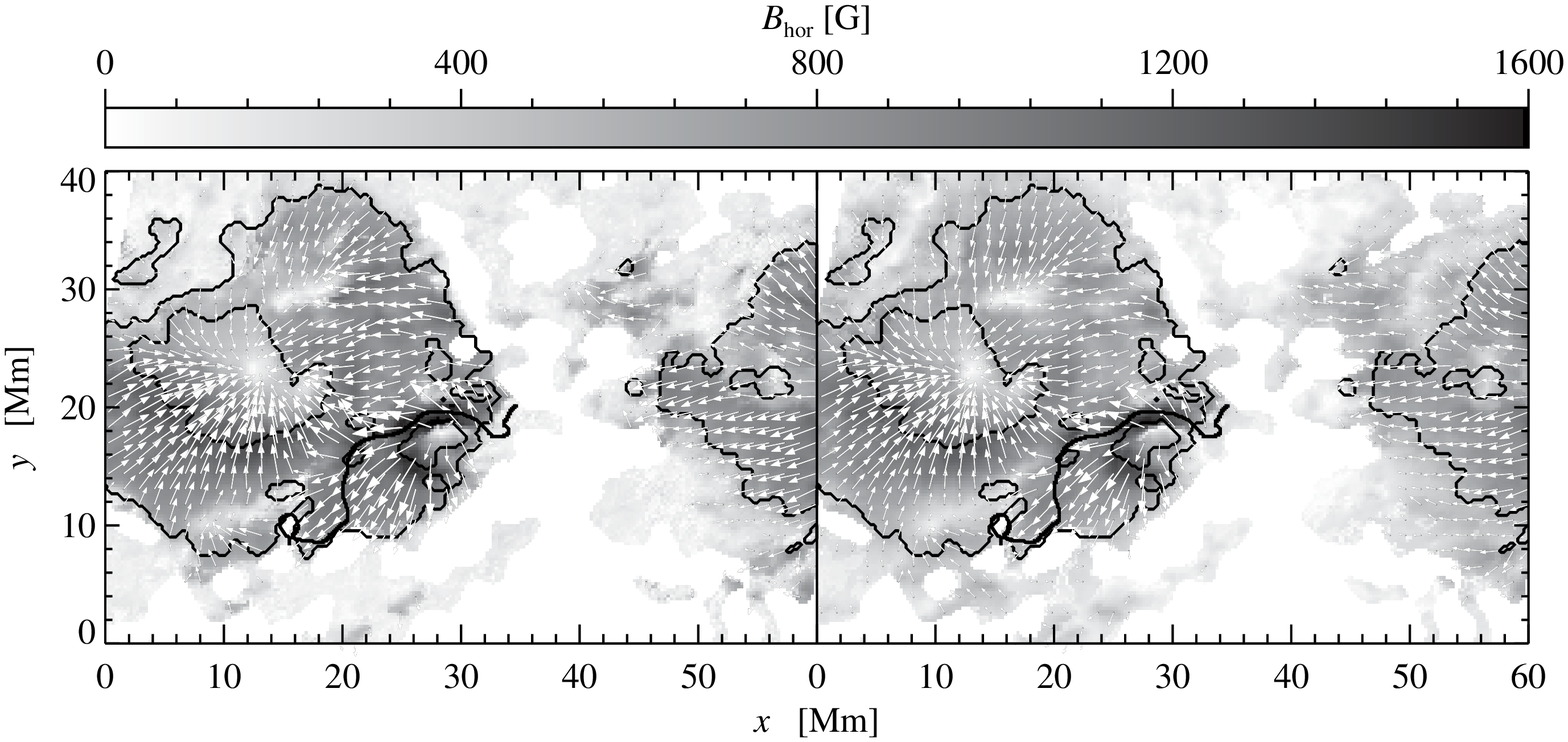}
\caption{Horizontal component of the magnetic field.
The strength is given by the gray-scale and the azimuth 
by the white arrows. 
}
\label{balthasar_bhor}
\end{center}
\end{figure}

The horizontal component of the magnetic field is shown in 
Fig.~\ref{balthasar_bhor}. We see a smooth transition from the main umbra
to the $\delta$-umbra. 
The strongest horizontal field surrounded the $\delta$-umbra. 
At some distance to the $\delta$-umbra,
close to a small bright feature in the intensity maps, the 
field lines were almost perpendicular to each other. At this location, 
the Stokes profiles were anomalous (see Fig.~\ref{balthasar_stokes}), 
and a single-component inversion was
not able to reproduce these profiles. \citet{balthasar-lites} interpreted
such a configuration by an interleaved system of magnetic field lines.

\subsection{Height dependence of the magnetic field}
\label{balthasar-subsec-height}

\begin{figure}[t]
\begin{center}
\includegraphics[width=132mm]{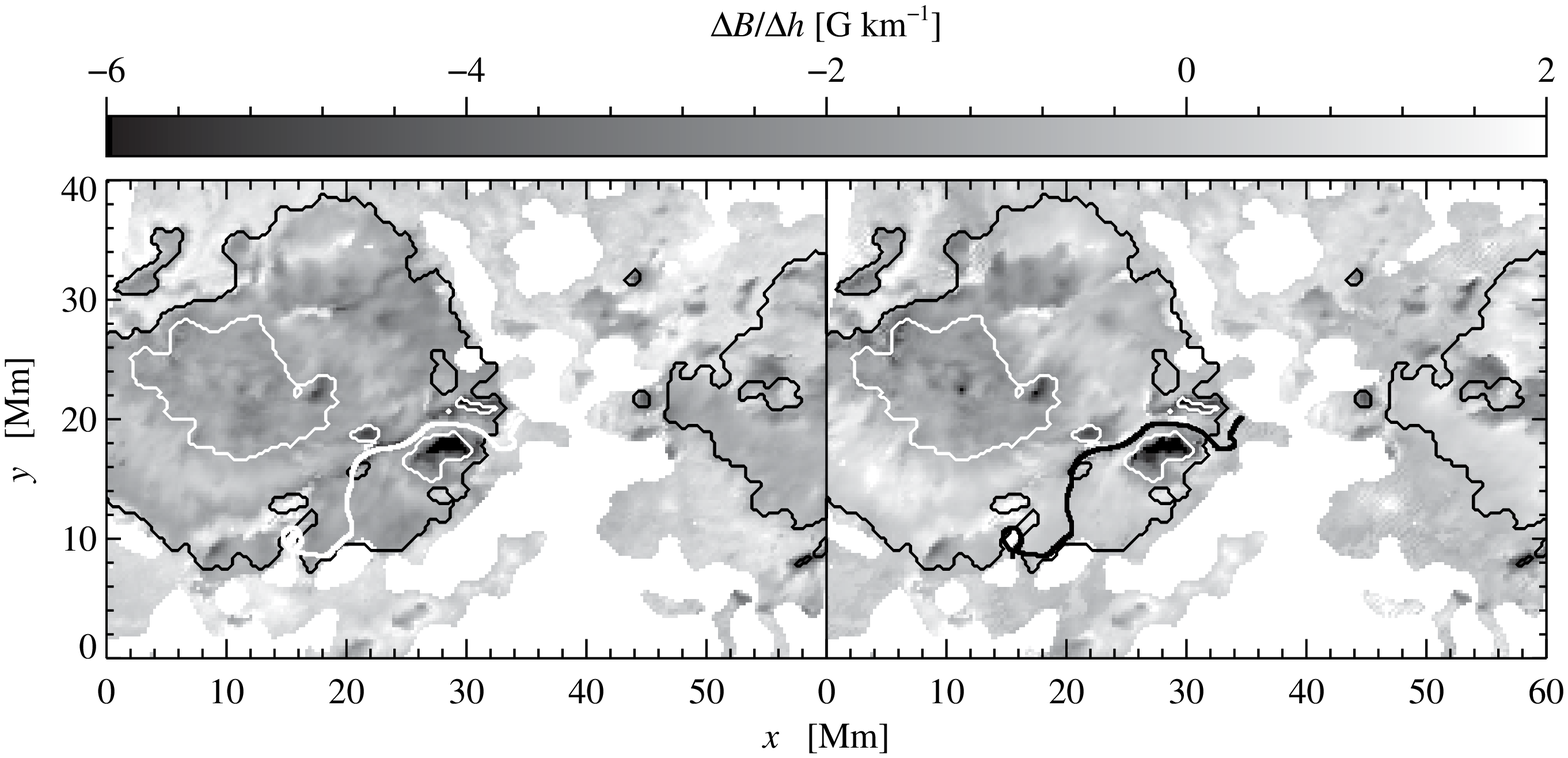}
\caption{Height gradients of the total magnetic field $B$ (left)
and the absolute value of the vertical component $B_{\mathrm z}$ (right) 
derived from the formation-height difference of the two lines
Fe\,{\sc i}\,1078.3\,nm and Si\,{\sc i}\,1078.6\,nm.
}
\label{balthasar_bdiff}
\end{center}
\end{figure}

Within the framework of our data we have two possibilities to determine 
the height dependence of the magnetic field. The first option is to take 
the difference of the magnetic quantities derived from the two lines and
divide it by the height difference for the two lines. The height differences 
were determined in the same way as by \citet{balthasar-hobagoem}, who used 
the depression contribution functions for two different model atmospheres
and interpolated according to the local temperature.
\textbf{The results are shown in Fig.~\ref{balthasar_bdiff}.}

In the main umbra, we found a mean decrease of the total magnetic field strength 
by 1.9\,G\,km$^{-1}$, which is comparable to values published for other spots. 
The decrease was much steeper in the $\delta$-umbra, here we encountered 
5.6\,G\,km$^{-1}$. Along the common part of PIL and CEL and in U3 the magnetic
field decreased by 3--4\,G\,km$^{-1}$, but along the CEL where it was separated 
from the PIL, the magnetic field strength was increasing by 1\,G\,km$^{-1}$.
A fast decrease was also observed in U4. Outside the spot, the magnetic 
field is increasing with height because of the canopy effect. 
The gradient of the absolute vertical component exhibited a similar behavior.
In the $\delta$-umbra, U3, and U4, the decrease was even slightly faster than for the 
total field strength indicating that the magnetic field became more horizontal
above these features. Along the whole CEL, $B_{\mathrm z}$ was increasing with height,
while the total field strength was decreasing in the common range of CEL and PIL. 
This discrepancy is explained by a fast decrease of the horizontal magnetic field
here.
In the mid penumbra, we observed that the vertical component increased 
with height because the magnetic field was more vertical in higher layers. 

The second option is applicable only for the vertical component of the 
magnetic field and starts from the condition that $\nabla \cdot \vec{B}$ is zero,
and vertical partial derivatives must be compensated by the horizontal ones:
 \begin{equation}
{\partial B_{\mathrm z} \over \partial z} = - \left({\partial B_{\mathrm x} 
\over \partial x} + 
{\partial B_{\mathrm y} \over \partial y} \right)
\label{balthasar-eq-divb}
\end{equation}
The horizontal derivatives were derived from differences of the values from
neighboring pixels, as described by \citet{balthasar-hoba06}.

\begin{figure}[!ht]
\begin{center}
\includegraphics[width=132mm]{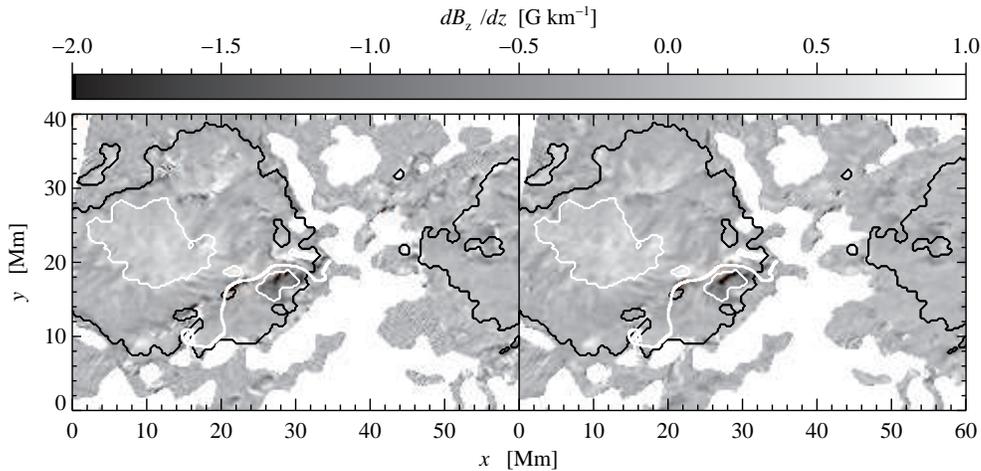}
\caption{Height gradients of the vertical component of the the magnetic field
derived via $\nabla \cdot \vec{B}=0$ from Fe\,{\sc i}\,1078.3\,nm (left)
and from Si\,{\sc i}\,1078.6\,nm (right). 
}
\label{balthasar_dbzdz}
\end{center}
\end{figure}

Qualitatively, the results were similar to those of the difference method,
but as \citet{balthasar-hobagoem} and \citet{balthasar-hvar} found, 
the values were smaller by a factor of about two\textbf{ as shown in 
Fig~\ref{balthasar_dbzdz}}. The vertical component of 
the magnetic field decreased faster with height in the $\delta$-umbra, U3 and U4 than
in the main umbra. Along the CEL, there was a tendency for an increasing $B_{\mathrm z}$
with height.


\subsection{Electric current densities}
\label{balthasar-subsec-jzhz}

\begin{figure}[t]
\begin{center}
\includegraphics[width=132mm]{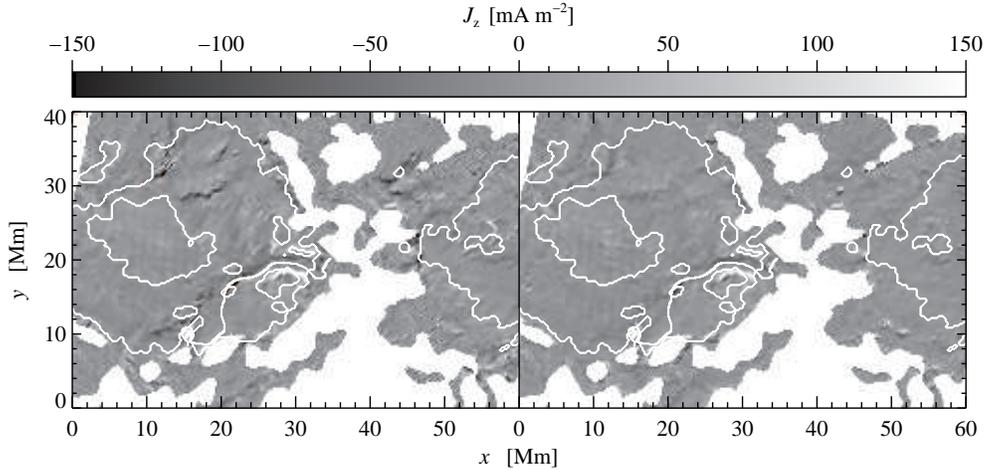}
\caption{Vertical component of electric current densities from 
Fe\,{\sc i}\,1078.3\,nm (left) and Si\,{\sc i}\,1078.6\,nm (right).
Values are clipped at $\pm$150\,mA\,m$^{-2}$.
}
\label{balthasar_strom}
\end{center}
\end{figure}

The horizontal partial derivatives of the magnetic field also allow us to
determine the vertical component of electric current densities $J_{\mathrm z}$
according to 
\begin{equation}
J_{\mathrm z} = {1 \over \mu}(\nabla \times \vec B)_{\mathrm z} =
{1 \over \mu}\left({\partial B_{\mathrm y} 
\over \partial x} - 
{\partial B_{\mathrm x} \over \partial y} \right),
\label{balthasar-eq-jz}
\end{equation}
where $\mu$ is the magnetic permeability. Partial derivatives were estimated from 
differences between neighboring pixels.
This procedure was used before by \citet{balthasar-hoba06}.
The results are shown in Fig.~\ref{balthasar_strom}. Positive current 
densities of more than 200\,mA\,m$^{-2}$ occurred near the outer boundary inside the $\delta$-umbra along a line parallel to the PIL. The maximum value from the iron line 
is 273 $\pm$ 76\,mA\,m$^{-2}$. Along the CEL, we detected negative current 
densities. Close to the $\delta$-umbra the negative current densities 
might be a counterpart to the positive values inside the $\delta$-umbra.
Negative current densities were still found in the part of the CEL, where it was 
separated from the PIL.
Strong current densities are much more pronounced in the 
Fe\,{\sc i}\,1078.3\,nm line than in the Si\,{\sc i}\,1078.6\,nm line, 
i.e., the currents occurred mainly in deep atmospheric layers.

\subsection{Doppler velocities and proper motions}
\label{balthasar-subsec-velo}

The Doppler velocities were investigated by \citet{balthasar-aaletter}.
The dominant feature in the photosphere was the Evershed-effect which 
was interrupted at the CEL. A special feature is a blueshift at the PIL
which \citet{balthasar-aaletter} interpreted as Evershed-effect related 
to the $\delta$-umbra.
In the  chromosphere above the CEL, small locations exhibited
downflows up to 8\,km\,s$^{-1}$, derived from the line 
Ca\,{\sc ii}\,854.2\,nm. One of these downflow patches was very close to 
the bright feature marked in Fig.~\ref{balthasar_kont}.
Another downflow patch is located just outside the penumbra, similar as a
patch observed by \citet{balthasar-hvar} near a single sunspot.
Other locations along the CEL exhibited upflows of the same order of magnitude 
as the downflows.

Proper motions also were investigated by \citet{balthasar-aaletter}. They used a 
time series of magnetograms from the Helioseismic and Magnetic Imager 
\citep[HMI,][]{balthasar-hmi}  
on board of the Solar Dynamic Observatory \citep[SDO,][]{balthasar-sdo} and applied the Differential 
Affine Velocity Estimator (DAVE) developed by \citet{balthasar-schuck2005,
balthasar-schuck2006}. This method delivered `magnetic flux transfer velocities'
\citep[][]{balthasar-schuck2006}, that do not necessarily represent plasma flows. 
The flux transfer velocities pointed towards the $\delta$-umbra, where its
Evershed effect was visible as blue-shift.
A flow of 0.25\,km\,s$^{-1}$ away from the $\delta$-umbra parallel
to the CEL was detected, which crossed the PIL, where it was separated from the 
CEL. On the other side of the CEL, towards the main umbra, only 0.05\,km\,s$^{-1}$
were found, resulting in a shear imbalance of 0.2\,km\,s$^{-1}$. 
A counterclockwise spiral motion covered a part of the $\delta$-umbra and 
U4, but we did not detect that the whole $\delta$-umbra was rotating.

\section{Discussion}
\label{balthasar-sec-discuss}

The discrepancy between different methods to derive the height gradient of 
the vertical component of the magnetic field strength is a long-lasting problem in 
solar physics and has been discussed by \citet{balthasar-lekamet},
\citet{balthasar-hobagoem}, and 
\citet{balthasar-hvar}. Determining geometrical heights from contribution or 
response functions has uncertainties, and one has to keep in mind that 
such functions cover an extended height range. Height gradients determined by this 
method but from different lines have similar values around 2\,G\,km$^{-1}$
\citep[see][] {balthasar-wittmann, balthasar-schmidt, balthasar-moran,
balthasar-lekamet}. To solve the problem
with larger differences of the line formation, one would need to extend the
solar atmosphere to much more than a few hundred kilometers. 
Similar gradients were also derived from height-dependent inversions as 
carried out by \citet{balthasar-westendorp}, \citet{balthasar-mathew}, and
\citet{balthasar-monica}.
However, comparing with a coronal C\,{\sc iv} line at 154.8\,nm, 
\citet{balthasar-hagyard} obtained gradients of 0.1--0.2\,G\,km$^{-1}$.
Using $\nabla \cdot \vec{B}$, \citet{balthasar-hofmann} obtained 0.32\,G\,km$^{-1}$ 
from data with low spatial resolution.
Many solar structures are rather small, i.e., at the 
spatial resolution limit of present instruments or even below. 
Thus, features that do not belong to the same 
magnetic structure enter the determination of the horizontal gradients and
affect the vertical gradients. 
Indications where found by 
\citet{balthasar-hvar} that higher spatial resolution decreases this discrepancy.
So far, the problem is not solved, but independent from the solution of this problem, 
we can state that the magnetic field decreases much faster with height above the
$\delta$-umbra than above the main umbra.

The CEL might be the dividing line between the original spot and new 
emerging flux forming the $\delta$-umbra. The penumbral area between PIL and CEL 
then would be the following part of the new bipolar flux. Penumbrae merged between 
main and $\delta$-umbra, i.e., between opposite polarities, similar as observed by 
\citet{balthasar-wang}, but not for the parts with the
same polarity. This scenario explains that the Evershed flow 
belonging to the main umbra ended at the CEL. The horizontal flux 
transfer velocities were parallel to the CEL and did not cross it, in contrast 
to the PIL. Strong changes of the magnetic field
with height in this area are not surprising, and there were electric currents at the CEL.
Such an emerging flux system would also 
explain that the CEL representing the dividing line between old and new flux is 
more important for the configuration of this $\delta$-spot than the PIL.

The observed chromospheric downflows above the CEL resemble the supersonic 
downflows found by \citet{balthasar-valentin} at the PIL of another sunspot,
but in the photosphere. We could not detect such large vertical velocities in the 
photosphere. Perhaps, the photospheric counterparts to the chromospheric flows 
were rather narrow, much less than the 2\arcsec{} in case of \citet{balthasar-valentin}, 
and they contributed only a small fraction of the signal in our resolution element.

\begin{figure}[t]
\begin{center}
\includegraphics[width=117mm]{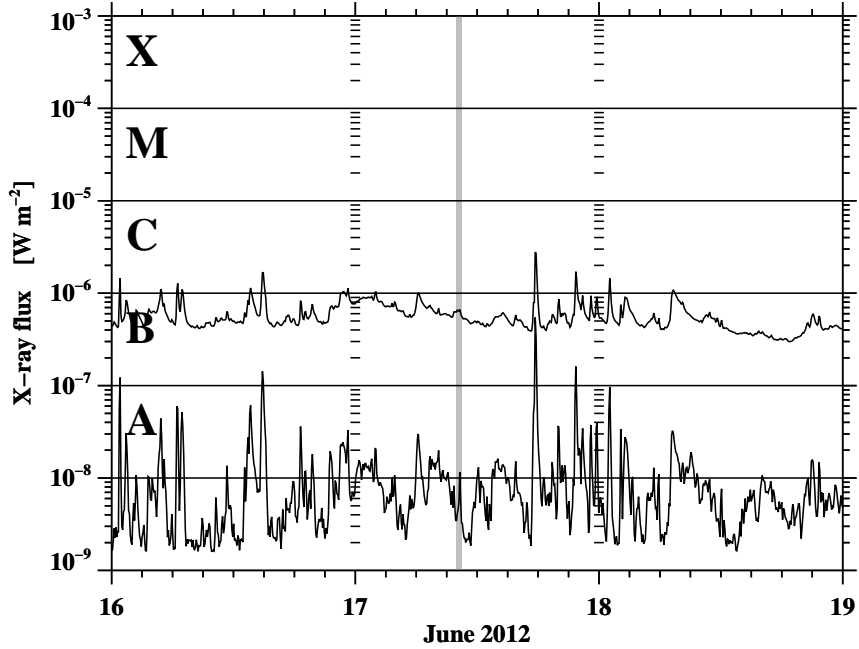}
\caption{X-ray flux in the 1.0--8.0~\AA\ (top) and 0.5--4.0~\AA\ (bottom) channels 
of the of the GOES-satellite for the period 2012, June 16--18.
The \textbf{gray} vertical bar marks our VTT-observing time. \textbf{For an easier 
readability, logarithmic scale marks are repeated at the beginning of each day.}
}
\label{balthasar-goes}
\end{center}
\end{figure}

The group NOAA 11504 produced a C1.8 flare about 19 hours before our observations
and a C3.9 flare seven hours after our observations as shown in Fig.~\ref{balthasar-goes}.
We did not find a shear flow in this group next to the $\delta$-umbra.
The flow away from it parallel to the CEL is probably not strong enough
to build up magnetic shear within a short period.
This indicates that a $\delta$-spot can be quiet in the sense of flares 
for at least a day.

\section{Conclusions}
\label{balthasar-sec-conclu}

In the following, we summarize the most important findings with regard to this 
$\delta$-spot in NOAA~11504.

\begin{description}
\item[$\bullet$] The magnetic field strength of 2250\,G in the $\delta$-umbra is somewhat 
        less than in the main umbra (2600\,G in deep photospheric layers).
\item[$\bullet$] We find that the magnetic field transition between the main
        and the $\delta$-umbra is rather smooth.
        A small location, where the magnetic azimuth changes by about 90$^\circ$
        from one pixel to the next, occurrs at some distance from the $\delta$-umbra 
        close to a bright patch next to the PIL.
\item[$\bullet$] The magnetic field decreases much faster with height above the $\delta$-umbra
        than above the main umbra.
\item[$\bullet$] Electric currents are detected at the CEL in deep photospheric layers
          and in the $\delta$-umbra.
\item[$\bullet$] Around the main umbra, we observe the typical Evershed flow which ends in the 
        southeastern part of the spot at the CEL. 
\item[$\bullet$] Related to the $\delta$-umbra, we detect a second system of Evershed flows. 
\item[$\bullet$] Large velocities of $\pm$8\,km\,s$^{-1}$ occur in the chromosphere 
          above the CEL.
\item[$\bullet$] No major flare was observed within seven hours before or after our observations.
\end{description}

We have shown that the CEL is more important than the PIL for this specific sunspot.
We are able to 
explain this assuming that the CEL is the dividing line between old 
and new emerging bipolar flux. 
Thus, $\delta$-spots can be stable, they do not always produce flares.
The brightenings in H$\alpha$ and the occurrence of central emission in the 
Ca\,{\sc ii}\,854.2\,nm line must be explained by physical processes 
related to the magnetic field in the chromosphere, and these processes play 
an important role for the configuration of this $\delta$-spot. Future observations 
should include also measurements of the chromospheric magnetic field.
This will be possible with, e.g., with the GREGOR Infrared Spectrograph 
\citep[][]{balthasar-gris} or the GREGOR Fabry P\'erot Interferometer 
\citep[][]{balthasar-gfpi} at the new GREGOR solar telescope \citep{balthasar-gregor}
in Tenerife.

\acknowledgments
The VTT and ChroTel are operated by the Kiepenheuer-Institut f\"ur Sonnenphysik 
(Germany) at the Spanish Observatorio del Teide of the Instituto de
Astrof\'\i sica de Canarias. The HMI-data have been used by courtesy of NASA/SDO 
and the HMI science team. MV expresses her gratitude for the
generous financial support by the German Academic Exchange Service (DAAD) in the 
form of a Ph.D. scholarship. CD and REL were  supported by grant DE 787/3-1 of the 
German Science Foundation (DFG).

\bibliography{balthasar-1-v2}

\end{document}